\documentclass{emulateapj}

\shorttitle{NICMOS Imaging of the W3 IRS 5 Cluster}
\shortauthors{Megeath et al.}

\bibpunct{(}{)}{,}{a}{}{;}
\bibpunct{(}{)}{,}{a}{}{;}

\begin{document}


\title{{\it Hubble Space Telescope} NICMOS Imaging of W3 IRS 5: A
  Trapezium in the Making?}


\author{S. T. Megeath\altaffilmark{1}, T. L. Wilson\altaffilmark{2,3},
M. R. Corbin\altaffilmark{4}}

\altaffiltext{1}{Harvard Smithsonian Center for Astrophyscis, MS-65, 60 Garden
St, Cambridge, MA 02138 (tmegeath@cfa.harvard.edu)}

\altaffiltext{2}{European Southern Observatory, K-Schwarzschild-Str. 2,
85748 Garching, Germany} 

\altaffiltext{3}{On leave from the Max Plack Institut f\"ur Radiastronomie, 69 Auf dem H\"ugel, Bonn, 53121 Bonn, Germany}

\altaffiltext{3}{Department of Physics and Astronomy, Arizona State University, Tempe AZ 85287}



\begin{abstract}

We present {\it Hubble Space Telescope} NICMOS imaging of W3 IRS 5, a
binary high-mass protostar.  In addition to the two protostars, NICMOS
images taken in the F222M and F160W filters show three new 2.22~$\mu$m
sources with very red colors; these sources fall within a region 5600
AU in diameter, and are coincident with a $\sim 100$~M$_{\odot}$,
dense molecular clump.  Two additional point sources are found within
$0.4''$ (800 AU) of one of the high-mass protostars; these may be
stellar companions or unresolved emission knots from an outflow.  We
propose that these sources constitute a nascent Trapezium system in
the center of the W3~IRS~5 cluster containing as many as five proto-OB
stars.  This would be the first identification of a Trapezium still
deeply embedded in its natal gas.

\end{abstract}



\section{Introduction}

In the pursuit of obtaining a comprehensive understanding of star
formation in our Galaxy, the formation of stars with masses greater
than 10~M$_{\odot}$ remains a challenging problem.  It has become well
established that massive stars form not in isolation, but in close
proximity to others stars spanning the entire range of the initial
mass function.  Observationally, the gregarious nature of massive
star formation has been established on three hierarchical levels. On
the largest spatial scales, young massive stars are almost always
found in clusters, with sizes around 0.5 pc \citep{hod94,sol04}.  The
more relevant scale for the formation of high-mass stars is that of a
few thousand AU.  On these spatial scales, massive stars are often
found to be part of multiple systems and Trapezia.  Trapezia, defined
as non-hierarchical (and consequently, unstable) multiple system, are
commonly found within OB-associations and open clusters
\citep{amb54,sha54}.  Multiplicity around high-mass stars continues to
even smaller separations; spectroscopic observations of massive stars
show companions with separations of 10 AU or less \citep{mer01}.

It is not clear whether OB stars form in these multiple systems, or
whether the multiple systems are formed from subsequent dynamical
interactions in the dense centers of young stellar clusters
\citep{bon98,kro01}.  To date, the known examples of OB stars in
multiple systems have already cleared their natal gas. If OB stars
form with companions at separations ranging from a few AU to 1000 AU,
then these companions may play important role in mediating accretion
onto OB stars by modifying accretion disks and envelopes, by adding
extra mass to the system \citep{ket03}, or through mergers
\citep{bon02}.

For this reason, it is essential to search for companions of forming
massive stars at the highest available angular resolution.  A rare
example of a binary high-mass protostellar system is W3 IRS 5, at a
distance of 1.8 kpc \citet{ima00}. This infrared source was first
resolved into a double infrared source by \citet{how81} and
\citet{neu82}. The total luminosity is $2 \times 10^5$~L$_{\odot}$
\citep{cam95}. Radio observations with the Very Large Array by
\citet{wil03} and \citet{tak04} show that each component of the
infrared source is coincident with radio continuum source with sizes
$< 240$~AU. These {\it hypercompact} HII regions are perhaps the
youngest detectable stage of a massive star. The W3~IRS~5 region is
also the source of at least two outflows
\citep{ima00,wil03}. \citet{meg96} found that the W3~IRS~5 cluster is
in the center of an embedded cluster of 80-240 low mass stars.  We
report here HST NICMOS observations of W3 IRS 5 with an angular
resolution of $350~AU$.

\section{Observations \& Data Reduction}

The observations were taken in Cycle 7 using camera 2 of NICMOS
onboard the {\it Hubble Space Telescope} \citep{tho98}.  A $3 \times
3$ map was made using a spiral-dither pattern with two dithers at each
map position.  The multiaccum mode was used with a total integration
time of 128~sec for each image.  The images were take in the F110W,
F160W and F222M filters.  The data were reduced with the standard
STSCI pipeline.  The frames were then registered and mosaiced using a
custom program written in IDL.

Photometry was obtained through a combination of aperture photometry
and point spread function (PSF) fitting.  The PSF was generated with
the TinyTim program \citep{kri97}. Using a custom IDL program, we fit
the PSF to five point sources in the F222M mosaic: NIR1, NIR2, NIR2a,
NIR2b and NIR4 (see Table~1).  For NIR2, NIR2a and NIR2b, which are
separated by less than $0.4''$, we fit three PSFs simultaneously.  We
also fit PSFs to five isolated point sources in the F222M mosaic and
measured the photometric offset between the PSF fitting and the
aperture photometry.  We used aperture photometry for the remaining
point sources in the F222M and F160W mosaics, including the more
isolated NIR3 and NIR5 sources.  The total count rates were measured
over a $3$ pixel aperture with a $5$ to $10$ pixel sky radius. To
determine magnitudes we used the equation $m = -2.5 \times log(PHOTFNU
\times CR / F_{\nu}(Vega))+Corr$, where CR is the count rate in
DN~sec$^{-1}$, $PHOTFNU = 2.077 \times 10^{-6}$ and
$5.502\times10^{-6}$~Jy~s~DN$^{-1}$, $F_{\nu}(Vega) = 1040.7$ and
$610.4$~Jy, and ``corr'', the aperture correction, is $-0.53$ and
$-0.71$ for the F160W and F222M bands, respectively.  The coefficients
are the Cycle 7 values taken from the online NICMOS data handbook, the
aperture corrections were measured from the data.  The resulting zero
points for the count rate were 19.40 and 21.22 for the F222M and F160W
bands.

\section{Results}

In the center of the W3 IRS~5 cluster is a small group of red objects,
the most prominent of which is the double near-IR source IRS~5
(Figures~1 and 2).  In Figure~1, we use lower limits to the colors of
sources undetected at F160W by adopting the $10 \sigma$ detection
limit of $20.56$~mag. Within a region less than $3''$ in diameter,
there are three sources with $[F160]-[F222] > 4$, and two additional
sources which have colors $[F160]-[F222] > 2$. We designate these
sources NIR $1-5$.  The five sources are coincident with a dense clump of
molecular gas, with a mass of 100~M$_{\odot}$ and FWHM diameter of
$5''$ or $9000~AU$ \citep{tie95,tie97,rob97}.  \citet{tie95} measure a
column density of $5.8 \times 10^{23}$~cm$^{-2}$ toward this region with
an $11''$ beam; the extinction through this clump in the F222M band is
60~magnitudes \citep{dra89}.  Consequently, we find it highly unlikely
that the detected stars are background stars.

Images of the W3 IRS 5 system in the F160W and F222M filter are shown
in Figure~3.  \citet{wil03} found that the two brightest infrared
sources, NIR1 and NIR2, are coincident with hypercompact HII regions
$D2$ and $B$. These were detected in the mid-IR by \citet{tak04}.
NIR3 was also detected in the radio continuum and mid-IR by
\citet{tak04}. To the east, there are two additional sources, NIR4 and
NIR5, which were not detected in the radio or mid-IR.  In panel C, we
show the same image with the two hypercompact HII regions subtracted
using a PSF generated with Tiny Tim \citep{kri97}.  The subtracted
image reveals two point sources near source NIR2; these are separated
from NIR2 by $0.3''$ and $0.4''$.  We designate these sources NIR2a and
NIR2b. In between NIR1 and NIR2 we find an elongated
nebulosity. 

In Table 1 we give the properties of the seven unresolved sources in
this region.  The coordinates were shifted so that NIR1 has the same
position as the $D2$ source in \citet{wil03}. Only two
sources were marginally detected in the F160M band, NIR1 and NIR4.
The measured magnitudes for NIR1 and NIR2 are 21.386 and 21.392, both
having 4.5 $\sigma$ detections.

\begin{table}
\begin{center}
\caption{Near-IR Sources \label{properties}}
\begin{tabular}{llllllll}
\tableline\tableline
  & R.A.   & Dec & F222M & Sep.$^{1}$ & P.A.$^{1}$  & Associated Radio  \\
  & (J2000) &  (J2000)   & Mag (unc) & A.U. & Deg. &or Mid-IR Source &  \\
1    & 2~25~40.784 & 62~05~52.62 & 12.21 (.03) & 2252 &  37 &  D2$^2$/Q5$^3$/K7$^3$/MIR1$^3$ \\
2    & 2~25~40.680 & 62~05~51.63 & 13.40 (.03) &  -   &   - &  B$^2$/Q3$^3$/K4$^3$/MIR2$^3$  \\
2a   & 2~25~40.690 & 62~05~51.91 & 14.78 (.07) &  519 &  14 & -  \\
2b   & 2~25~40.682 & 62~05~51.23 & 16.38 (.07) &  730 & 177 & - \\
3    & 2~25~40.727 & 62~05~49.93 & 17.94 (.10) & 3176 & 169 & Q4$^3$/K6$^3$/MIR3$^3$ \\
4    & 2~25~40.896 & 62~05~51.71 & 16.20 (.07) & 2780 &  87 & -  \\
5    & 2~25~41.040 & 62~05~52.10 & 17.94 (.10) & 4712 &  80 & - \\
\tableline
\tablenotetext{1}{Separation (Sep.) and Position Angle (P.A.) measured relative
to NIR2.  Separations assume a distance of 1830 pc \citep{ima00}}
\tablenotetext{2}{Associated 2 cm radio source  using nomenclature of \citet{cla94}, also detected at 1.3 and 0.7~cm by \citet{wil03}}
\tablenotetext{3}{Associated 1.3~cm, 0.7~cm or mid-IR source using nomneclature of \citet{tak04}}
\end{tabular}
\end{center}
\end{table}

The detection of NIR1, NIR2 and NIR3 at radio and mid-IR wavelengths
essentially confirms that these are young OB stars embedded in the
molecular core. NIR4 does not have a detected HII region, suggesting
that either the emission is quenched by a high accretion rate
\citep{ket03}, that the star is not hot enough to produce significant
Lyman continuum radiation, or that NIR4 is a background star - which
is unlikely given the high extinction.  Using the the tentative F160W
detection, NIR4 has a color of $(F160W-F222M) = 5.2$.  Assuming that
the emission from NIR4 is from a hot photosphere with no contribution
from circumstellar dust, implying an intrinsic color of $(F160W-F222M)
= 0$, and adopting a reddening law of $A_\lambda/E(J-H) = 2.4 \times
\lambda^{(-1.75)}$ \citep{dra89} and a distance modulus of 11.23, we
derive an absolute F222M magnitude of $M_{222} = -1.7$.  Using the
tabulation of $K$-band magnitudes vs mass for ZAMS OB stars in
\citet{han97}, this magnitude corresponds to a spectral type B0.5V and
a mass of 11~$M_{\odot}$.  The lack of radio emission suggests that
the star has a later spectral type than B0.5, and it is likely that
NIR4 is a later type star undergoing pre-main sequence evolution. For
example, the Herbig Ae/Be star BD+65~1637, which has a spectal type of
B3 and a mass of 9.5~M$_{\odot}$, has a dereddened absolute $K$-band
magnitude of $-1.5$, similar to the of NIR4 \citep{hil92}.  The Lyman
continuum flux of a B3 star is 3\% that of a B1 star
\citep{pan73}. For NIR5 there is only a lower limit to the measured
color, $(F160W-F222M) > 2.7$.  Assuming an intrinsic color of
$(F160W-F222M) = 0.025$ (i.e. equal to the $H-K$ color with an M3
dwarf \citet{bes88}), the resulting absolute magnitudes in the F222M
band is $< 3.2$, corresponding to masses $> 0.13$ and $>
0.3$~M$_{\odot}$ for ages of 0.1~Myr and 1~Myr respectively
\citep{mue02}. This source is probably a low mass star.

The two remaining point sources are found within a projected distance
of 800 AU of NIR2; given their proximity to NIR2 and their red colors,
we find it likely that these are stellar companions to NIR2.  Neither
of these sources would be resolved in the mid-IR imaging of
\citet{tak04}, but they would be resolved in VLA radio observations of
this region. We see no radio counterpart to the fainter source, NIR2b,
this source has the same magnitude as NIR5 and is probably a low mass
star.  NIR2a is directly between the radio sources $B$ and $A$.  If
the 2.22~$\mu$m emission from NIR2a is pure photospheric emission from
an embedded star, the star would have to be quite massive; the upper
limit of the dereddened F222M magnitude is equivalent to that of a
ZAMS O8.5 star \citep{han97}. An alternative explanation is that the two
sources are unresolved knots of shocked H$_2$ from a bipolar flow
originating from NIR2. \citet{wil03} argued that NIR2/radio source $B$
was driving an outflow in a roughly north-south direction; as shown in
Figure 3, the two features are relatively aligned with the flow.
These two explanations can be tested through narrow-band images in the
H$_2$ $1 \rightarrow 0$ S(1) line.

We also report the first detection of a compact nebula between the
NIR1 and NIR2. The nebula may result from the collision of outflows
from NIR1 and NIR2, or from the collision of an outflow from one source
with the envelope of the other.  \citet{wil03} and \citet{tak04} find
evidence for a ionized knot moving from NIR1 at 100~km~s$^{-1}$ which
may be part of an outflow moving to the north-east. If the proposed
outflow is bipolar and centered on NIR1, the counter-flow would be
moving to the south-west in the direction of NIR2.  Observations of
\citet{cla94} showed a radio source between NIR1 and NIR2, $D1$,
which was not detected in subsequent VLA imaging.  \citet{wil03} argue
that these transient radio sources are created in outflows, in this
case $D1$ may tracing the outflow responsible for the observed near-IR
nebula. We speculate that the X-rays detected by \citet{hof02} toward
W3~IRS5 may originate in this nebula.

\subsection{Discussion}

Could W3~IRS~5 be a precursor to a Trapezium still embedded in its
natal gas?  Trapezia are defined as non-hierarchical multiple system
of three or more stars; \citet{abt00} proposed that as a working rule
for identifying Trapezia, the largest projected separation should be
no more than three times the smallest projected separation.  The
separations between the five near-IR sources ranges between 5600 AU
and 1800 AU and roughly satisfy this criteria.  At smaller
separations, the detection of candidate companions around NIR2
provides the first evidence for multiplicity at separations $< 1000$
AU around high-mass protostars.  Companions with separations ranging
from 460~AU to 16 AU have been detected toward the OB stars composing
the Orion Nebula Trapezium system \citep{wei99}.  Hence, the spatial
distribution of the primaries, and the presence of companions is
similar to that observed in other Trapezium system.

The maximum projected separation of the five near-IR sources is 5600
AU, compared to 10000 AU for the Trapezium in the Orion Nebula.
\citet{abt00} identified fourteen likely Trapezia, the median radius
to the furthest outlying member of these Trapezia is 40000 AU. Hence,
the projected separations in the W3~IRS~5 system are smaller than
those observed in optically visible Trapezia.  The stellar surface
density of the five near-IR sources composing the W3~IRS~5 Trapezia is
10000~pc$^{-2}$, four times the density in the surrounding cluster -
which is 2600~pc$^{-2}$ for a cluster radius of $21''$ \citep{meg96}.

The small size of the W3~IRS~5 system may be a function of its youth;
such systems may expand as the natal gas is expelled from the system
by outflows and radiation.  Maps of the molecular gas show a compact
core of dense gas surrounding the W3 IRS 5 with a mass of $\sim
100$~$M_{\odot}$ \citep{tie95, tie98, rob97}.  The Lyman-continuum
fluxes required by the observed radio continuum fluxes imply spectral
types of B0.5 for the stars associated with NIR1, NIR2 and NIR4
\citep{wil03,tak04,vac96}. Since the radio continuum measurements
suffer from a high optical depth, this is a lower limit.  A B0.5 star
has a mass around 18~M$_{\odot}$ \citep{vac96}, this gives a lower
limit to the combined stellar mass of 54~M$_{\odot}$.  \citet{meg96}
found that the star formation efficiency, i.e. the ratio to the
stellar mass to the total stellar and gas mass is about $18\%$ in the
W3~IRS~5 cluster.  However, in the IRS~5 system, the stellar mass
appears to be $> 30\%$ of the total mass. This high star formation
efficiency may allow Trapezia to survive the dispersal of the natal
gas and emerge as bound systems within a larger cluster.

\clearpage

\begin{figure}
\epsscale{1.}
\plotone{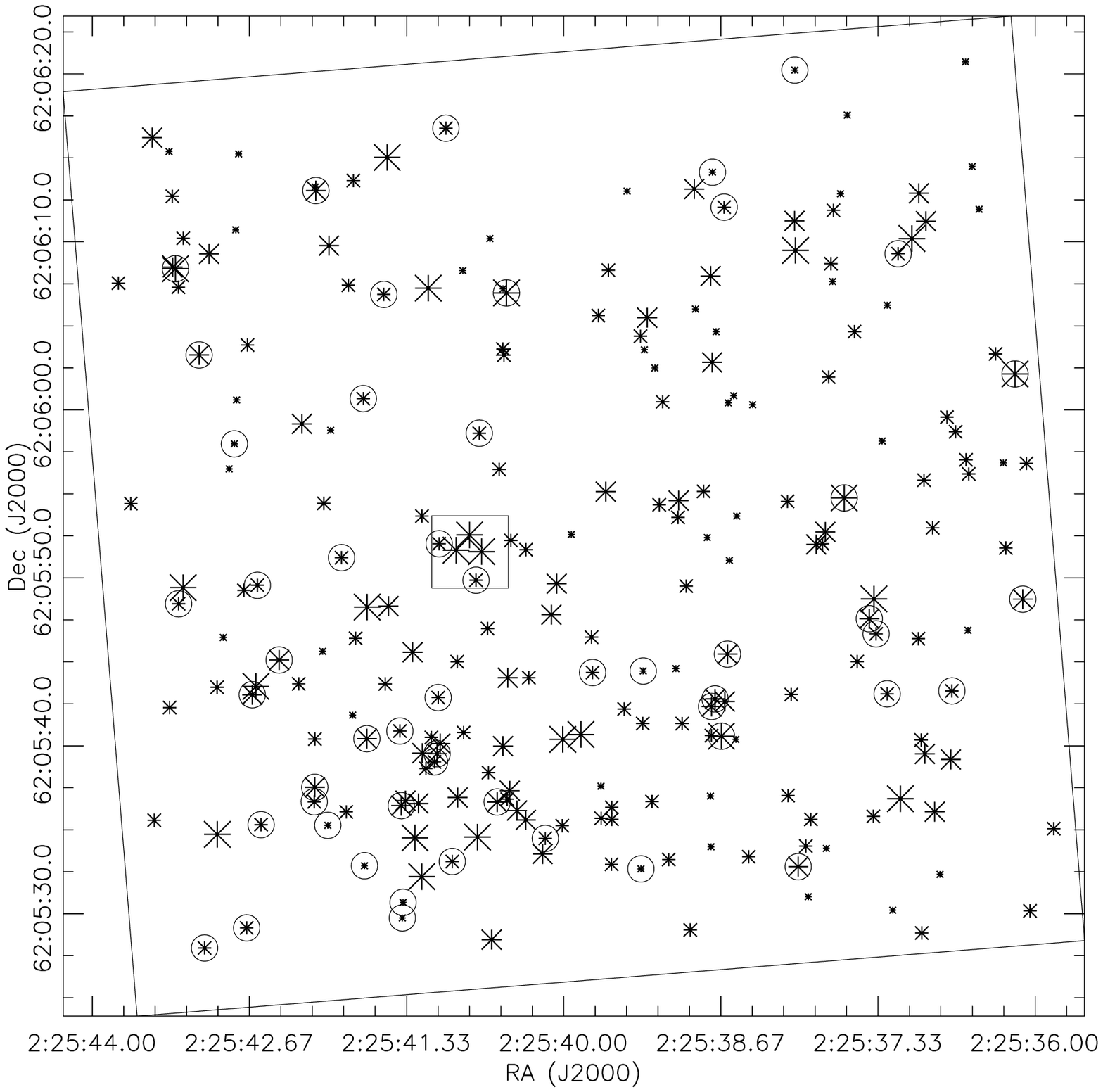}
\caption{The distribution of red sources detected in the F222M band.
  The size of the asterisk depends on the color of the source, ranging
  from $1 < (F160W - F222M) \le 2$ for the smallest asterisks, $2 <
  (F160W - F222M) \le 3$, $3 < (F160W - F222M) \le 4$ and $(F160W -
  F222M) > 4$ for the largest asterisks.  For sources undetected in
  the F160W band, a magnitude of 20.56, equivalent to a 10~$\sigma$
  detection, is given as an lower limit.  Sources with only lower
  limits to their colors are marked by circles; their colors may be
  redder than indicated by the size of the asterisk.  The outer box
  shows the extent of the field imaged with NICMOS, the inner square
  is the region displayed in Figure 3.}
\end{figure}

\begin{figure}
\epsscale{1.}
\plotone{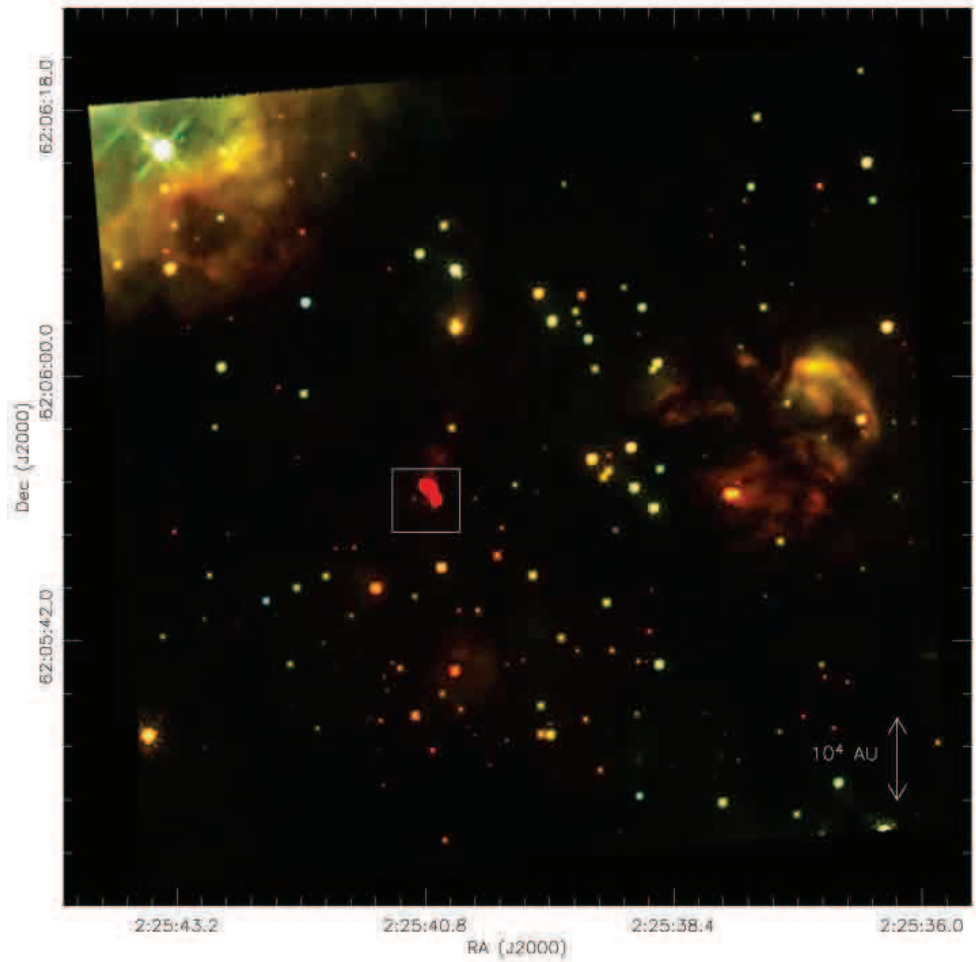}
\caption{A color composite image constructed from the F110W (blue),
  F160W (green) and F222M (red) mosaics of the W3~IRS~5 region,
  encompassing the whole region surveyed in the NICMOS
  measurements. The box shows the region displayed in Figure 3.}
\end{figure}

\begin{figure}
\epsscale{1.}
\plotone{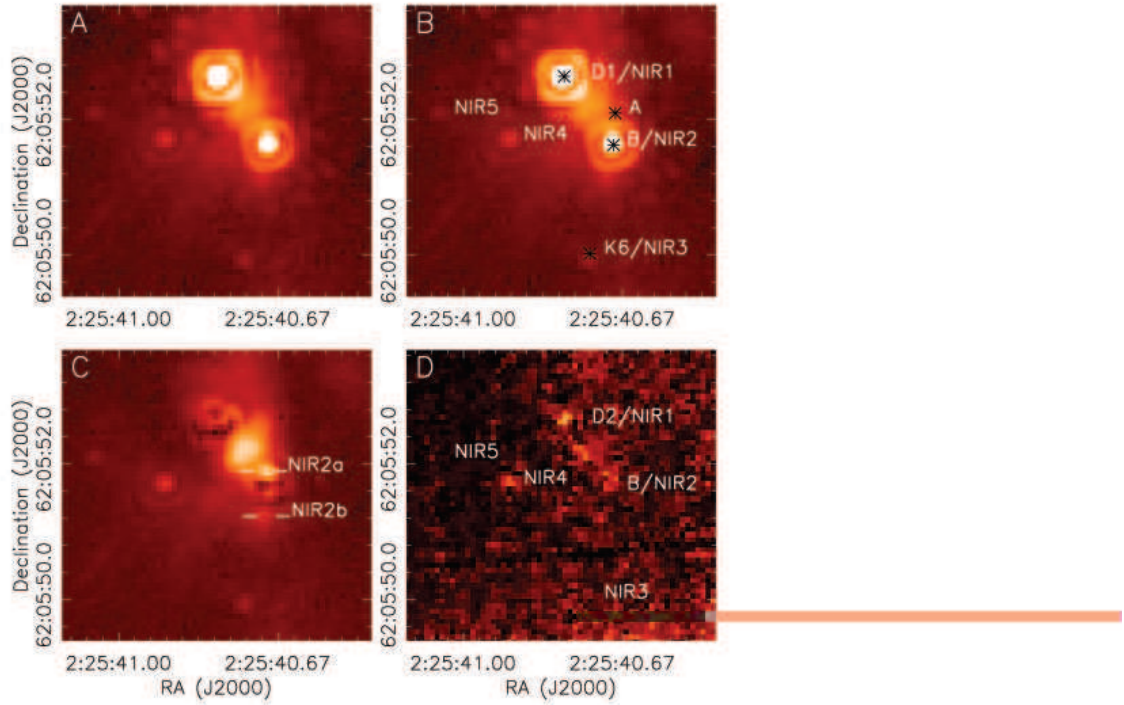}
\caption{F222M ($2.22$~$\mu$m) and F160W ($1.60$~$\mu$m) images of W3
  IRS 5 and the neighboring red sources and nebulosities.  In Panel A
  we show the F222M image using a cube root scaling.  In Panel B we
  show the same image, with the main NIR sources marked.  The 
  asterisks mark the positions of the associated radio sources $D2$,
  $B$, $A$ and $K6$.  In Panel C we show the image with the NIR1 and
  NIR2 sources subtracted.  An extended nebulosity between the two
  sources is clearly evident.  Two additional point sources sources
  partially hidden by the PSF of NIR2 are marked. The ring--like
  pattern is a residual from the PSF subtraction. In Panel D we show
  the F160W image toward this region, with the five IR sources marked.}
\end{figure}

\end{document}